\documentclass[lettersize,journal]{IEEEtran}
\usepackage{amsmath,amsfonts}
\usepackage{algorithmic}
\usepackage{algorithm}
\usepackage{array}
\usepackage[caption=false,font=normalsize,labelfont=sf,textfont=sf]{subfig}
\usepackage{textcomp}
\usepackage{stfloats}
\usepackage{url}
\usepackage{verbatim}
\usepackage{graphicx}
\usepackage{cite}
\usepackage{upgreek}
\usepackage{multirow}
\usepackage{multicol}
\usepackage{comment}
\usepackage{caption}
\usepackage{tikz}
\renewcommand\fbox{\fcolorbox{red}{white}}
\newcommand\submittedtext{%
  \footnotesize This work has been submitted to the IEEE for possible publication. Copyright may be transferred without notice, after which this version may no longer be accessible.}

\newcommand\submittednotice{%
\begin{tikzpicture}[remember picture,overlay]
\node[anchor=south,yshift=10pt] at (current page.south) {\fbox{\parbox{\dimexpr0.65\textwidth-\fboxsep-\fboxrule\relax}{\submittedtext}}};
\end{tikzpicture}%
}
\newcommand\copyrighttext{%
  \footnotesize \textcopyright \the\year{} IEEE. Personal use of this material is permitted. Permission from IEEE must be obtained for all other uses, including reprinting/republishing this material for advertising or promotional purposes, collecting new collected works for resale or redistribution to servers or lists, or reuse of any copyrighted component of this work in other works.}

\begin{document}

\title{AR-SFQ: Asynchronous Reset Library Using $\alpha$-Cell Design}

\author{\IEEEauthorblockN{Yasemin Kopur, Beyza Zeynep Ucpinar, Mustafa Altay Karamuftuoglu, Sasan Razmkhah, Massoud Pedram,~\IEEEmembership{Fellow,~IEEE,}}
\thanks{This work has been funded by the National Science Foundation (NSF) under the project Expedition: Discover (Design and Integration of Superconducting Computation for Ventures beyond Exascale Realization) grant number 2124453.}
\thanks{Y. Kopur, B. Z. Ucpinar, M. A. Karamuftuoglu, S. Razmkhah, and M. Pedram are with Ming Hsieh Department of Electrical and Computer Engineering, University of Southern California, Los Angeles, USA.} \thanks{e-mails: razmkhah@usc.edu, pedram@usc.edu}}

\markboth{Transaction on Applied Superconductivity}%
{Shell \MakeLowercase{\textit{et al.}}: A Sample Article Using IEEEtran.cls for IEEE Journals}


\maketitle

\submittednotice

\begin{abstract}
Rapid Single Flux Quantum (RSFQ) circuits are the most evolved superconductor logic family. However, the need to clock each cell and the deep pipeline causes a complex clock network with a large skew. This results in lower throughput and high latency in RSFQ. This work introduces an asynchronous RSFQ cell library that incorporates the $\alpha$-cell, enabling bidirectional signal paths in RSFQ circuits. The $\alpha$-cell mitigates the need for a large clock network by allowing reverse signal flow, minimizing routing, and enabling compact circuit designs. We demonstrate the library's reliability and efficiency by analog simulations and using in-house optimization tools. The asynchronous reset RSFQ (AR-SFQ) will enable efficient implementation of scalable, high-performance computing frameworks, such as state machines, neuromorphic computing, and higher fan-in circuits.
\end{abstract}

\begin{IEEEkeywords}
single flux quantum, bidirectional transmission cell, asynchronous reset, superconductor electronics
\end{IEEEkeywords}

\section{Introduction}
\IEEEPARstart{S}{uperconductor} electronics (SCE), particularly SFQ circuits, offer significant advantages in clock speed and power consumption over traditional CMOS technology. By representing data through quantized magnetic flux, SFQ circuits achieve ultrafast switching speeds and low power dissipation, addressing the growing need for efficient and high-performance computing \cite{likharevRSFQ}. In contrast, CMOS, which has long dominated the field, faces increasing power and speed limitations as Moore's law reaches its end \cite{badaroglu2021more}. SFQ technology provides a promising solution to these challenges, positioning itself as a strong candidate for next-generation computing architectures \cite{razmkhah2023superconducting}. However, this technology poses its unique challenges.

In RSFQ technology, data is encoded using the presence or absence of a single flux quantum (SFQ) pulse, where the presence of a pulse signifies a logic '1.' Its absence represents a logic' 0.' Consequently, the distinction between a logic '0' and no data is established by the presence of a clock pulse, making RSFQ logic inherently pipelined. This encoding approach is fundamental to managing data flow and storage in superconducting electronics (SCE). Unlike traditional circuits, advanced SFQ technologies can incorporate pulse direction as an additional dimension in data encoding. By leveraging pulse direction, SFQ architectures can support a broader range of applications and significantly enhance processing capabilities and overall performance in SCE.

The complex clock network, coupled with the fanin/fanout of one in RSFQ circuits, makes complex logic synthesis impractical. Several studies have been done to examine the wiring cells \cite{mixed_wiring} and interconnects to achieve efficient cell placement and routing \cite{interconnnect_routing, placement_routing_jtlptl}. The timing analysis of the wiring cells has been explored in \cite{stat_timing, timing_char_rsfq} in more detail. Despite the inherent capabilities of RSFQ technology for bidirectional computation, current interconnect designs and logic designs do not support this functionality. For instance, the splitter (SPL) and confluence buffer (CBU), also known as Merger, interconnect cells are dedicated to unidirectional operations, which restricts their ability to facilitate bidirectional data flow. To address this limitation, we demonstrate a new interconnect $\alpha$-cell in \cite{alpha_cell}. This novel design enables bidirectional data propagation in SFQ circuits. The $\alpha$-cell utilizes bidirectional I/O ports, effectively preserving the core interconnect functionality while enhancing operational flexibility. This feature not only offers alternative implementations for circuits but also significantly improves the operational capabilities of RSFQ systems.

To overcome challenges in large-scale integration of SCE, the creation of efficient cell libraries becomes crucial for high-performance, functional, and scalable design. Such libraries are instrumental in boosting design capabilities by offering optimized building blocks to meet the needs of modern applications. Several works demonstrate these cell libraries, showing significant strides in both RSFQ \cite{coldflux_cell_library, haolin_alljj, maezawa2004design, rsfq_cell_library_maezawa} and ERSFQ \cite{rsfq_ersfq_library} technologies. In \cite{razmkhah2024high, cong2024superconductor}, we show new cell libraries for inductorless SFQ logic by using novel devices. These studies highlight the need for robust RSFQ and ERSFQ systems to achieve efficiency and adaptability in superconductor electronic circuits.

In this work, we extend the capabilities of RSFQ circuits by integrating $\alpha$-cells with basic logic gates and memory configurations to add asynchronous reset to the SFQ cells, which are called AR-SFQ. AR-SFQ can be used to incorporate complex functional designs without the burden of a complex clock network. Integrating a reset functionality requires additional peripheral components, which increases the number of Josephson junctions (JJs) and consequently diminishes the overall yield. This increased complexity makes circuits more vulnerable to process variations. However, by leveraging $\alpha$-cell, these challenges can be effectively mitigated. Additionally, the propagation of pulses through an $\alpha$-cell to the output of a logic gate or memory cell can modify its state and decrease the internally stored magnetic flux by multiples of the flux quantum, thereby enhancing pulse management in state machines. As a result, AR-SFQ is capable of supporting the implementation of state machines and neuromorphic applications while enhancing fan-in and enabling the generation of complex functions, as explained throughout this paper.

The key contributions of this paper are as follows:
\begin{itemize}
\item Mitigation of circuit complexity by extending SFQ circuit capabilities via state reset
\item Increased fan-in with different functional implementations by datapath management
\end{itemize} 

In the next section, we will discuss the methodology used for designing the asynchronous SFQ logic using $\alpha$-cells. In the next chapter, we will discuss the memory design methodology and how we can implement a more compact memory with this logic. In section IV, we will discuss the architectural benefits of this implementation, and the last chapter is the conclusion.

\section{Asynchronous design methodology}
\subsection{$\alpha$-cell}
The $\alpha$-cell is a novel interconnect cell designed for bi-directional data propagation, combining the functionalities of a Confluent Buffer Unit (CBU) and a Josephson Transmission Line (JTL). It features three inputs and two outputs, with its structure optimized for propagating SFQ pulses. The cell's behavior is governed by asymmetric JJ loops, which control the flow of SFQ pulses.

This asymmetry allows the $\alpha$-cell to function as a JTL when signals travel from input to output and as a CBU, performing parallel-to-serial data conversion when signals arrive from inputs 2 or 3. The $\alpha$-cell adds flexibility by performing dual roles and is compatible with the RSFQ cell library. The wiring cell has been validated in the \cite{alpha_cell}, demonstrating that the cell supports bi-directional data transfer. The cell offers multiple functionalities, some of which will be explored in detail in this paper. The optimization analysis for every cell is done by using qCS (quantum Cell Studio) as the algorithm showed in \cite{margin_analysis}.

\subsection{AR-SFQ Logic Gates}
Logic cells play a crucial role in circuit design by enabling fundamental operations such as AND, OR, XOR, and inversion. The $\alpha$-cell enhances these functionalities by providing bi-directional data propagation and enabling the construction of more complex and efficient logic circuits. By incorporating the $\alpha$-cell, we can achieve versatile logic blocks that reduce the need for additional components while improving performance.

\subsubsection{AND Gate}
\begin{figure}[htb!]
\centering
\begin{minipage}{\linewidth}
    \centering
    \includegraphics[width=1\textwidth]{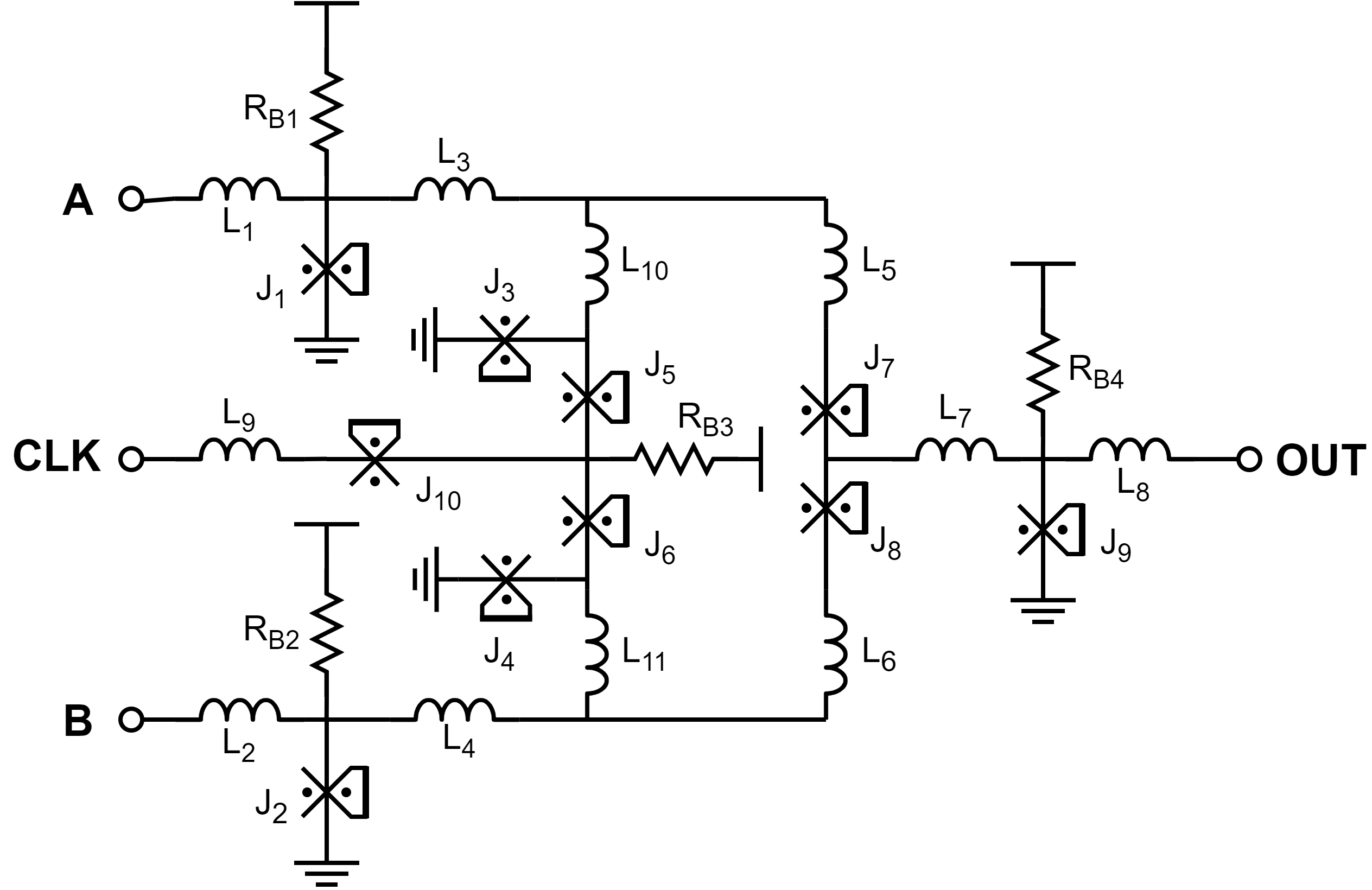}
    \caption{\small Schematic of the AND gate. Configuration: L1=2.1 pH, L2=2.1 pH, L3=5.57 pH, L4=5.58 pH, L5=4.04 pH, L6=2.1 pH, L7=7.58 pH, L8=7.58 pH, L9=4.1 pH, J1=148.82 $\mu$A, J2=112.32 $\mu$A, J3=117.86 $\mu$A, J4=113.91 $\mu$A, J5=111.55 $\mu$A, J6=78.11 $\mu$A, J7=90.46 $\mu$A, J8=96.99 $\mu$A, J9=169.68 $\mu$A, J10=102.3 $\mu$A}
    \label{fig:and_sch}
\end{minipage}

\begin{minipage}{\linewidth}
    \centering
    \includegraphics[width=1\textwidth]{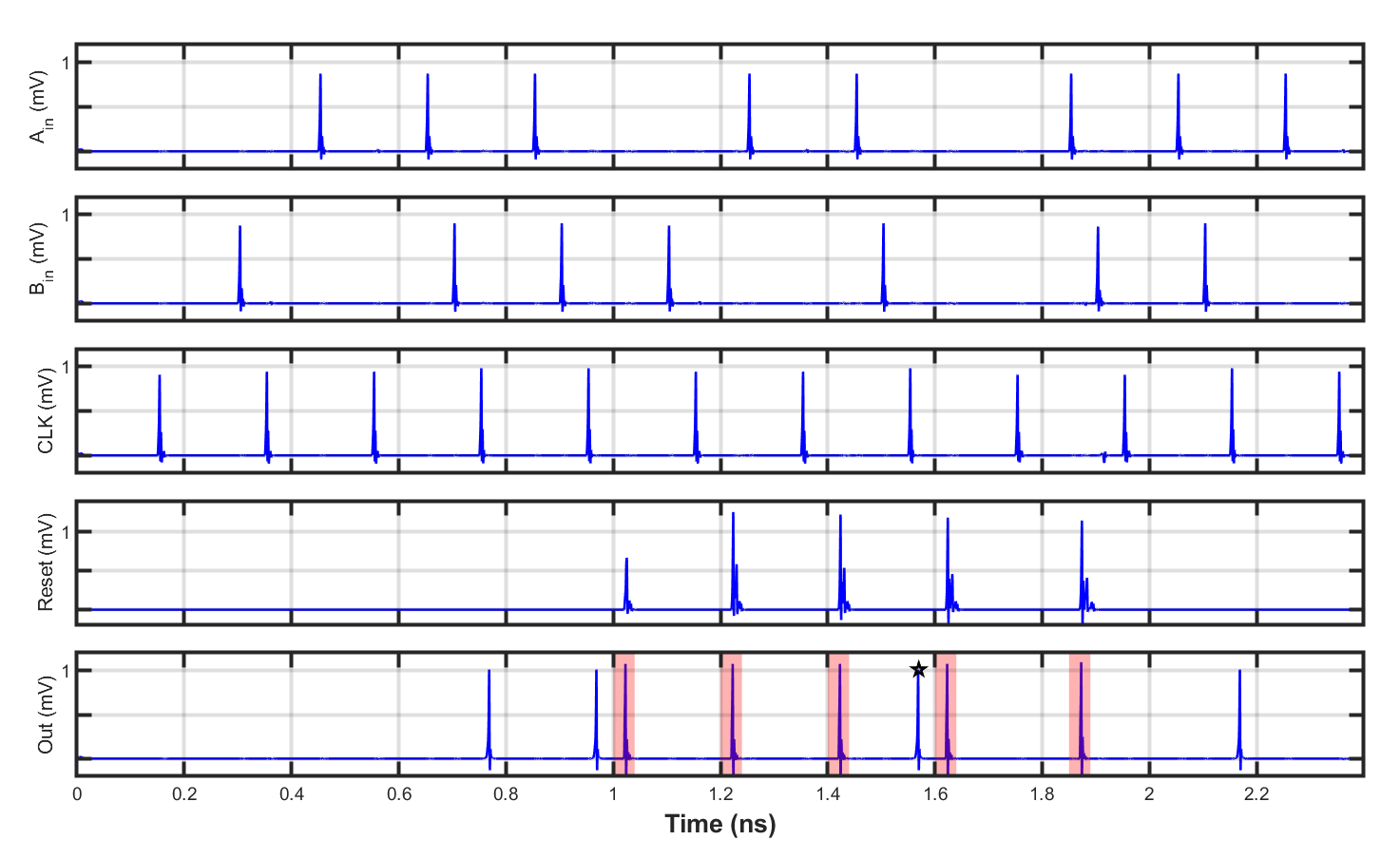}
    \caption{Simulation of the AND gate.}
    \label{fig:and_sim}
\end{minipage}
\end{figure}

The RSFQ implementation of the AND gate can be described as a configuration of two destructive readout structures that establish a delicate equilibrium among four escape junctions. As illustrated in Fig.~\ref{fig:and_sch}, L7 and L8 establish storing the data in both directions. In Fig.~\ref{fig:and_sch}, J5 and J6 function as primary escape junctions, which respond to the clock (CLK) input when no data is stored within the loop. Upon receiving a logic-1 from either input, the corresponding junction is deactivated by the current flowing through that input's path. J7 and J8 serve as secondary escape junctions, which may remain active unless both inputs are logic-1. In this balanced condition, where both inputs are logic-1, J7 and J8 are deactivated as well, allowing the output pulse to overflow through the output junction J9.
A reset signal, delivered in the reverse direction, reactivates the primary escape junctions, restoring the circuit to its default (0,0) state. This reset mechanism allows the circuit to handle subsequent inputs in the same clock cycle. For instance, if the loop receives another logic (1,1) after resetting and during the same clock cycle, it will overflow again through J9, behaving in the same manner as it usually does. This behavior can be attributed to the absence of inductive elements in the reverse direction, preventing it from retaining state information in the reverse direction. As demonstrated in the simulation results shown in Fig.~\ref{fig:and_sim}, the gate is capable of producing a logic-1 output when the reset is triggered before both input signals are applied.
The component values are listed in Fig.~\ref{fig:and_sch}. The critical margin for this gate is [-26,22]\% for J9, the output junction. The AND gate requires the reset signal to be applied within a defined timing window. The reset must be issued no earlier than 6.84 ps before the arrival of the first input signal and no later than 4.5 ps before the clock pulse (setup time) to ensure proper operation. These timing limitations can be mitigated through a phased approach to signal management. By ensuring that input signals are received within the first 80\% of the clock cycle and the reset signal is consistently applied at the end of the clock cycle, the risk of timing conflicts and associated errors can be effectively minimized.

\subsubsection{OR Gate}
\begin{figure}[htb!]
\centering
\begin{minipage}{\linewidth}
    \includegraphics[width=1\textwidth]{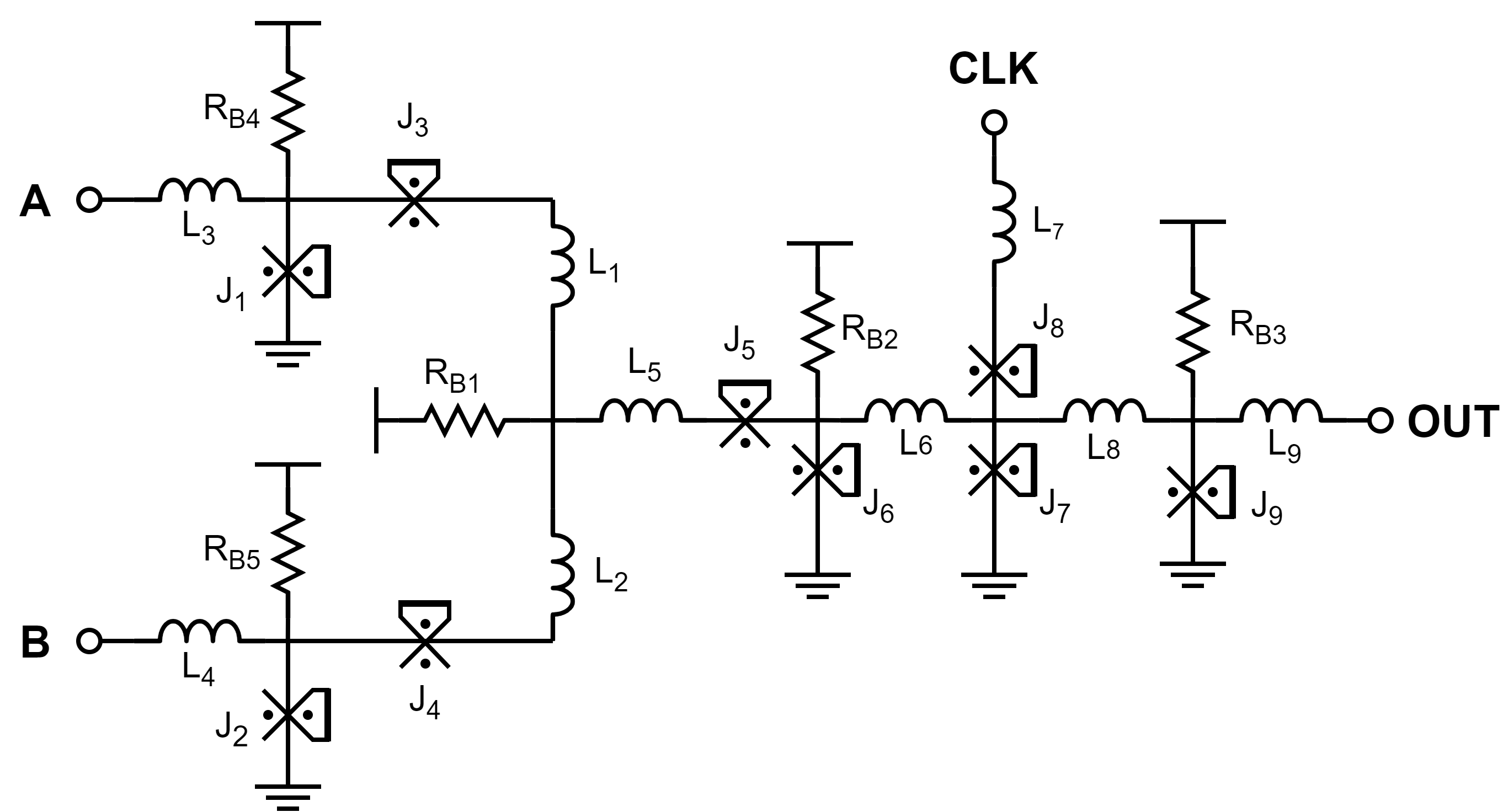}
    \centering
    \captionof{figure}{\small Schematic of the or gate. Configuration: L1=1.58 pH, L2=1.36 pH, L3=1.94 pH, L4=2.19 pH, L5=5.22 pH, L6=9.27 pH, L7=1.45 pH, L8=9.09 pH, L9=1.38 pH, J1=163.46 $\mu$A, J2=178.41 $\mu$A, J3=109.23 $\mu$A, J4=113.13 $\mu$A, J5=94.038 $\mu$A, J6=106.62 $\mu$A, J7=118.01 $\mu$A, J8=91.04 $\mu$A, J9=123.86 $\mu$A}
    \centering
    \label{fig:or_sch}
\end{minipage}

\vspace{3 mm}

\begin{minipage}{\linewidth}
    \includegraphics[width=1\textwidth]{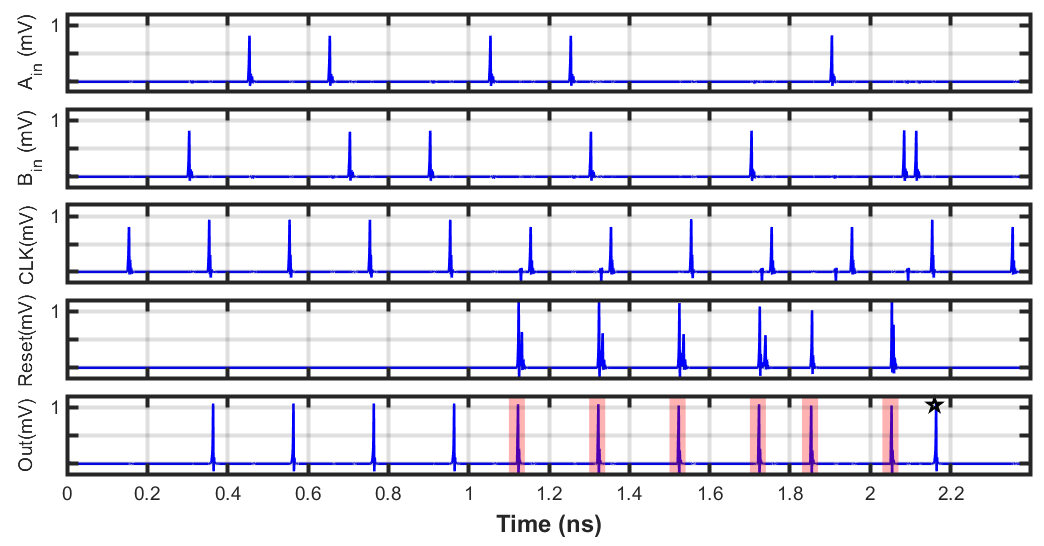}
    \centering
    \captionof{figure}{Simulation of the or gate.}
    \centering
    \label{fig:or_sim}
\end{minipage}%
\end{figure}

The RSFQ implementation of the OR gate can be defined as a destructive readout (DRO) connected to two different inputs by a merging stage. There are some improvements to avoid the backpropagation in the merger stage. In Fig.~\ref{fig:or_sch}, the merger stage is supported by escape junctions J3 and J4. While the structure is similar, there are minor variations in the component values, which are initially tuned manually and then optimized using the QCS algorithm. 
A reset signal received from the output pin resets the data stored in the loop but does not function as memory in the reverse direction. Specifically, if there is no stored flux at the DRO stage, a reset signal will not generate a reverse pulse at the input port. However, a reset signal can arrive before the input data, and it can suppress the first input signal received within the same clock cycle. If a premature reset occurs and two input pulses are received during the same clock cycle, the output will still register as logic-1. Both scenarios are illustrated in the simulation results in Fig.~\ref{fig:or_sim}. 
The finalized values are presented in Fig.~\ref{fig:or_sch}, with a critical margin of [-29,50]\%, primarily influenced by L6 and RB2, which are the elements of the storing inductance loop. The setup time of the reset signal is 7ps before the CLK signal. For a single signal window, there is no time dependency between the input and the reset signals to capture the functionality. However, if a multiple pulse are received, reset must arrive at least 5.39 ps after the latest signal. If double pulses occur for the same input within 2.3 ps, they are treated as a single pulse. Similarly, if pulses from different inputs arrive within 4.68 ps, they are considered a single event, and no reset timing limitations apply.

\subsubsection{XOR Gate}
\begin{figure}[htb!]
\centering
\begin{minipage}{\linewidth}
    \includegraphics[width=1\textwidth]{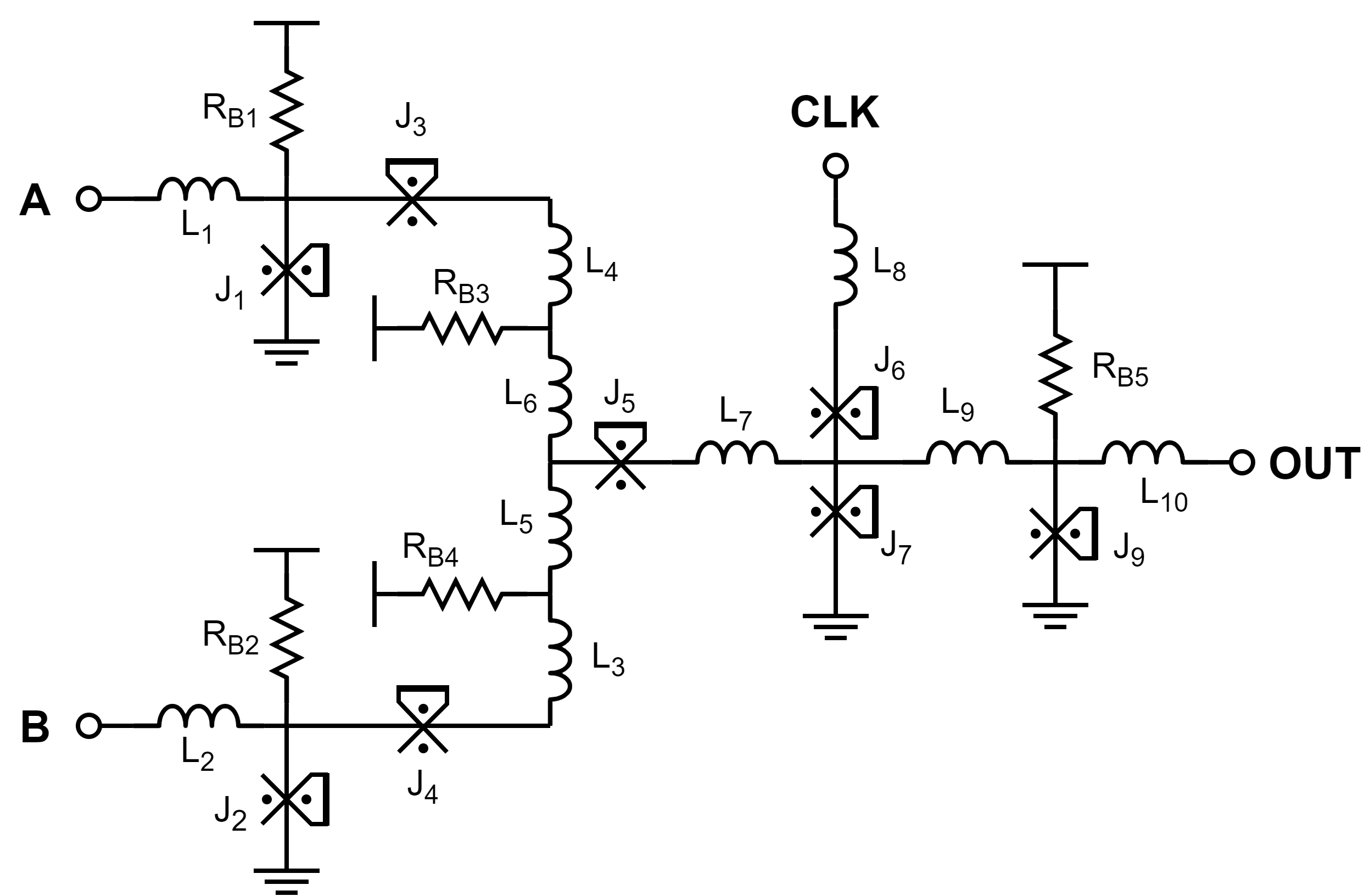}
    \centering
    \captionof{figure}{\small Schematic of the xor gate. Configuration: L1=2.53 pH, L2=1.91 pH, L3=1.68 pH, L4=1.12 pH, L5=8.02 pH, L6=9.08 pH, L7=1.45 pH, L8=4.44 pH, L9=7.54 pH, L10=2.56 pH, J1=203.71 $\mu$A, J2=127.03 $\mu$A, J3=112.49 $\mu$A, J4=114.48 $\mu$A, J5=116.04 $\mu$A, J6=93.49 $\mu$A, J7=111.18 $\mu$A, J8=168.81 $\mu$A}
    \centering
    \label{fig:xor_sch}
\end{minipage}

\vspace{2 mm}

\begin{minipage}{\linewidth}
    \includegraphics[width=1\textwidth]{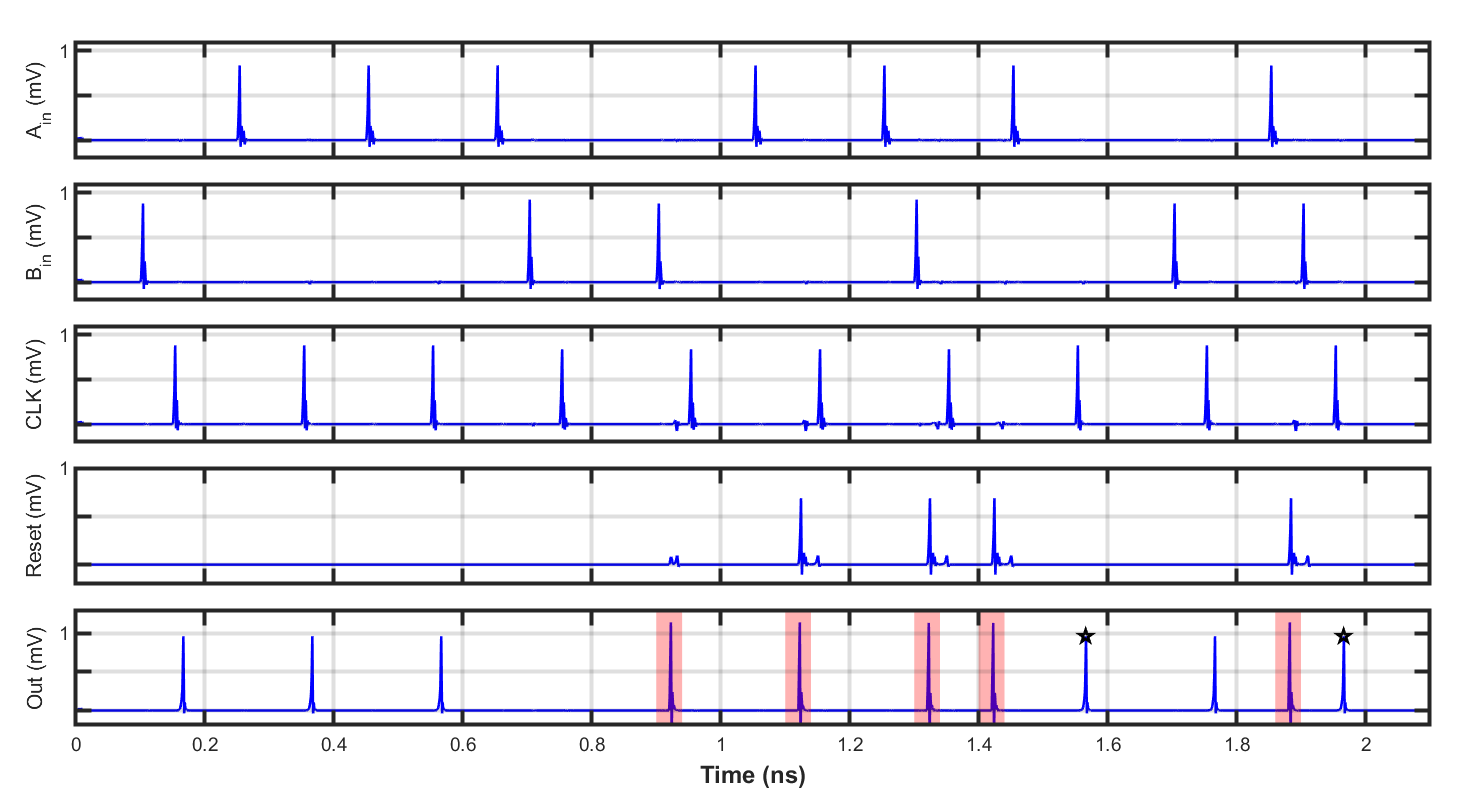}
    \centering
    \captionof{figure}{Simulation of the xor cell.}
    \centering
    \label{fig:xor_sim}
\end{minipage}%
\end{figure}

The XOR gate, shown in Fig.~\ref{fig:xor_sch}, features two competing storage loops, which tend to overflow immediately through the escape junction J5. As the two reverse loops are interconnected, the structure resets the flux stored in the first loop when the second input receives a logic-1. The stage before the clock (CLK), where L7 is connected, contains a primary flux loop designed to store the dominant data from the competing loops. Since J5 has a higher threshold than typical escape junctions, it does not switch when only one logic-1 is present, allowing the primary loop to transmit the stored value to the output upon receiving the CLK pulse.
A reset signal applied from the output direction resets both loop stages to their default state. The reverse signal similarly affects the data loop when two logic-1 inputs are received, resetting each other. However, instead of J5 switching in this scenario, one of the active escape junctions dampens the signal, as the flux of the reset signal is equivalent to that of a standard input. Similar to the AND gate, the data-storing loop features an escape junction, and the reset lacks memory. This makes the circuit vulnerable to signals received later in the same clock cycle. An example of this error is illustrated in Fig.~\ref{fig:xor_sim}, with a star marking the peak.
The component values are listed in Fig.~\ref{fig:xor_sch}. The critical margin for this gate is [-26,35] \% and [-28,31] \%  for J5 and J6, respectively. While J5 is the mentioned escape junction for the logic(1,1) damping mechanism, J6 is the clock escape junction. The XOR gate also requires the reset signal to be applied within a defined timing window. The reset must be issued no earlier than 15.14 ps before the arrival of the first input signal and no later than 10 ps before the clock pulse (setup time) to ensure proper operation.

\subsubsection{Inverter Gate}
The asynchronous RSFQ inverter is a recent addition to our library, offering a functional interpretation of the reset option as an inverter. Unlike other gates in the library, which utilize an alpha gate to reset the circuit between clock signals, this inverter employs the reset pin as its only input. While the alpha gate is still used to provide a negative flux input, its purpose here is not to reset the circuit but rather to enable the inverter functionality.

The operation integrates with a multi-flux DRO (MDRO) gate, where the positive input of the MDRO is connected to the clock signal. This connection introduces a flux value of +1 with each clock pulse. If no negative flux input is provided, the MDRO outputs a single SFQ pulse, representing logic 1. However, if a negative flux input is applied via the alpha gate, the +1 and -1 fluxes cancel, resulting in zero flux and no output, interpreted as logic 0.

Although this inverter does not eliminate the need for a reset signal, its design reuses the reset mechanism for functional operation. This approach offers potential advantages, such as increased compactness compared to a standard RSFQ inverter and a higher error margin.

\begin{figure}[htb!]
\centering
\noindent\begin{minipage}{\linewidth}
    \includegraphics[width=0.65\textwidth]{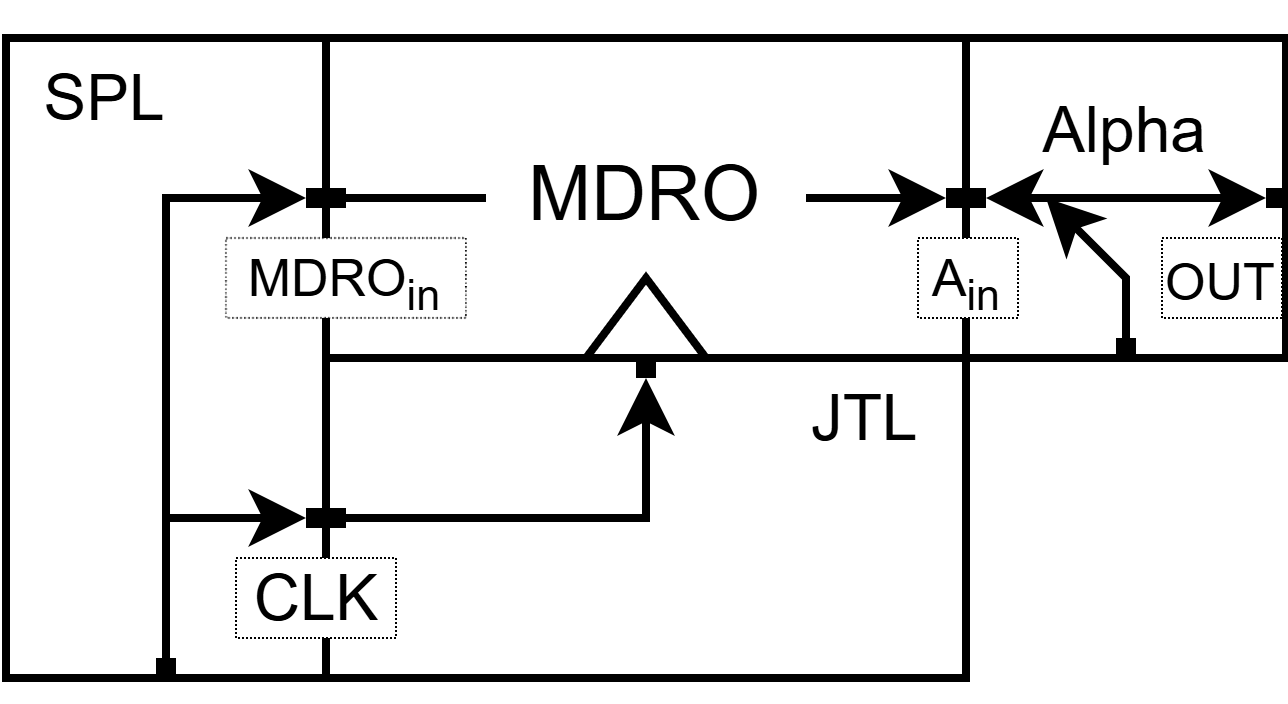}
    \centering
    \captionof{figure}{Schematic of the inverter gate.}
    \centering
    \label{fig:inv_sch}
\end{minipage}%

\vspace{2 mm}

\begin{minipage}{\linewidth}
    \includegraphics[width=1\textwidth]{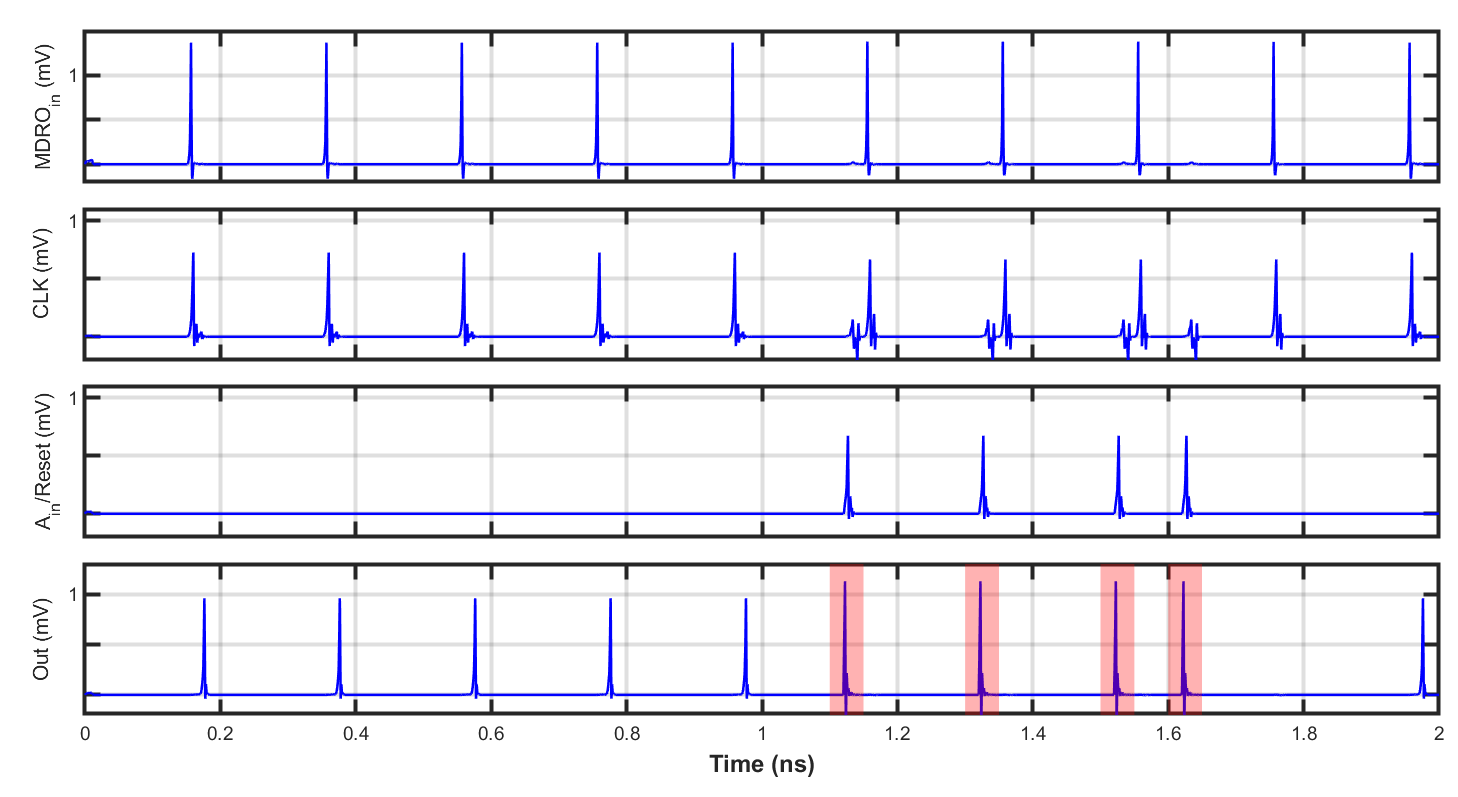}
    \centering
    \captionof{figure}{Simulation of the inverter cell.}
    \centering
    \label{fig:inverter_sim}

\end{minipage}%
\end{figure}

The inverter schematic is shown in Fig.\ref{fig:inv_sch}, with simulation results in Fig.\ref{fig:inverter_sim}, highlighting the circuit's behavior under different input conditions. The MDRO retains its original circuit parameters, as detailed in Fig.~\ref{fig:mdro_sch}..

This design showcases an innovative application of RSFQ principles, emphasizing functional simplicity while maintaining compatibility with existing asynchronous logic structures.

\section{$\alpha$-Memory design}
Memory cells are essential components in circuit design, enabling data storage and supporting functions like toggling, counting, and state transitions. The $\alpha$-cell enhances these capabilities by offering bi-directional data transfer and efficient state manipulation within memory systems. By integrating the $\alpha$-cell, we can achieve more adaptable memory blocks by setting initial states, including reset operations.

\subsection{Delay Flip Flop (DFF)}
A D flip-flop (DFF) is a fundamental memory unit in digital electronics that stores a single bit of data. It captures the value of its data input (D) at the moment of a clock signal transition, typically on the rising edge, and holds this value stable at its output (Q) until the next clock cycle. This ability to retain data over time makes DFFs essential in applications such as registers, counters, and state machines, where they ensure reliable data storage and synchronization in sequential logic circuits. 

\begin{figure}[htb!]
\centering
\noindent\begin{minipage}{\linewidth}
    \includegraphics[width=0.7\textwidth]{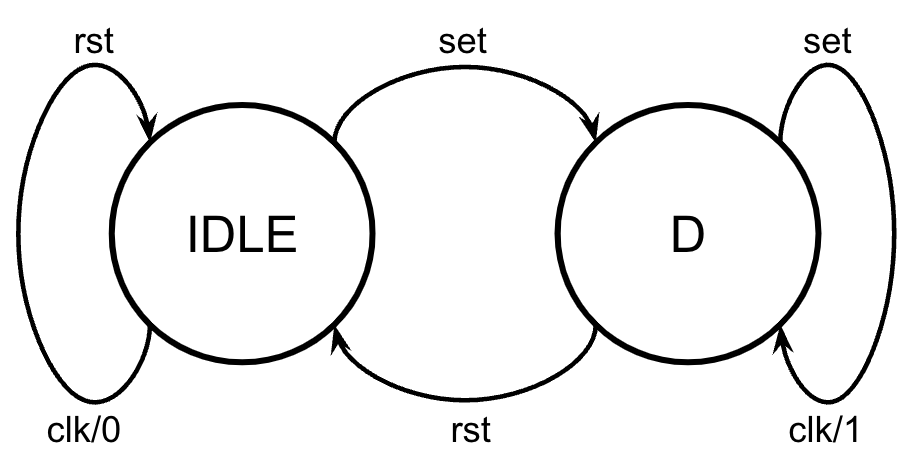}
    \centering
    \captionof{figure}{State machine of the dff gate.}
    \centering
    \label{fig:dff_sm}
    
    \includegraphics[width=0.8\textwidth]{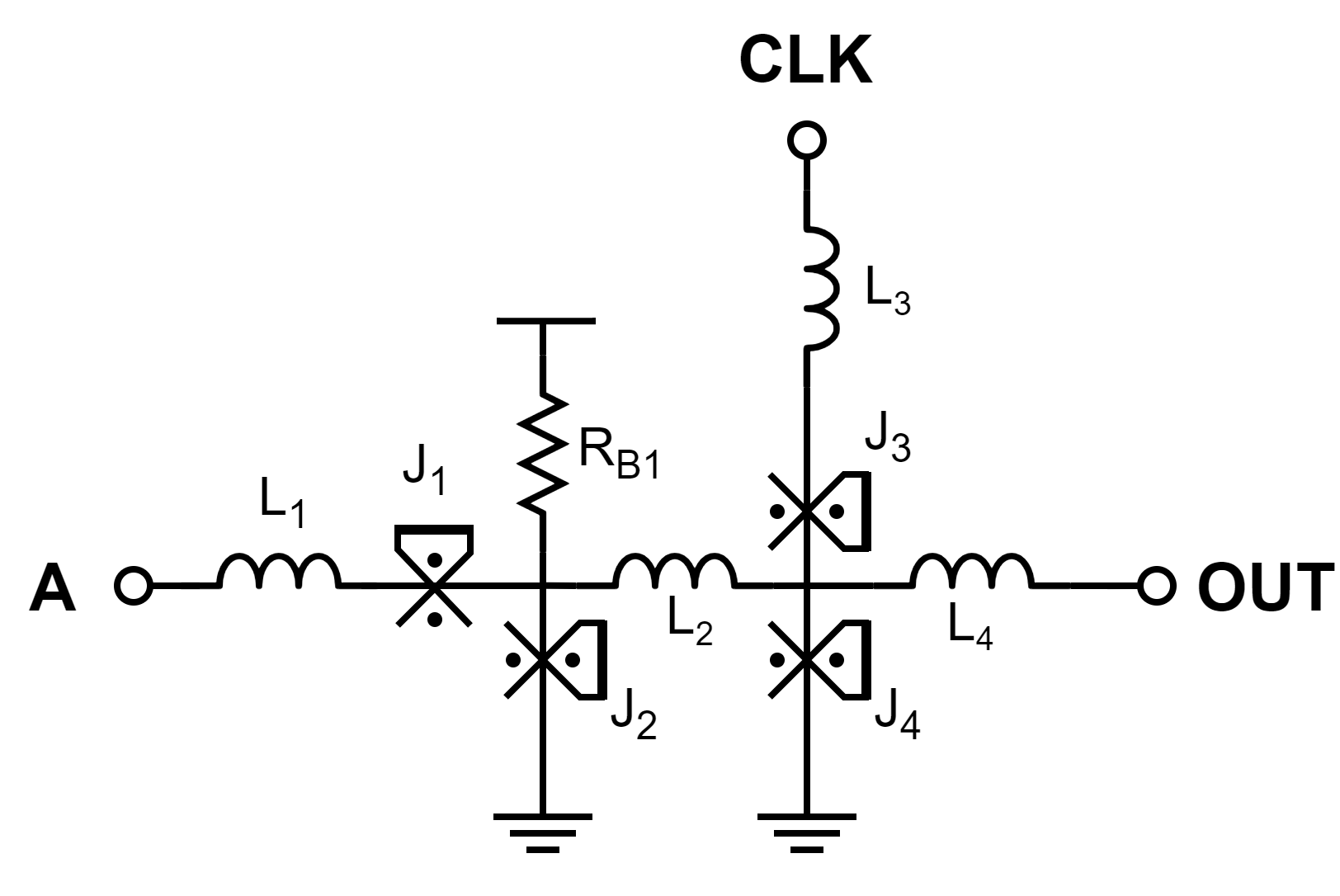}
    \centering
    \captionof{figure}{\small Schematic of the dff gate. Configuration: L1=2.10 pH, L2=6.67 pH, L3=4.45 pH, L4=1.55 pH, J1=208.52 $\mu$A, J2=201.78 $\mu$A, J3=180.00 $\mu$A, J4=239.29 $\mu$A}
    \centering
    \label{fig:dff_sch}
\end{minipage}

 \vspace{3 mm}
 
\begin{minipage}{\linewidth}
    \includegraphics[width=1\textwidth]{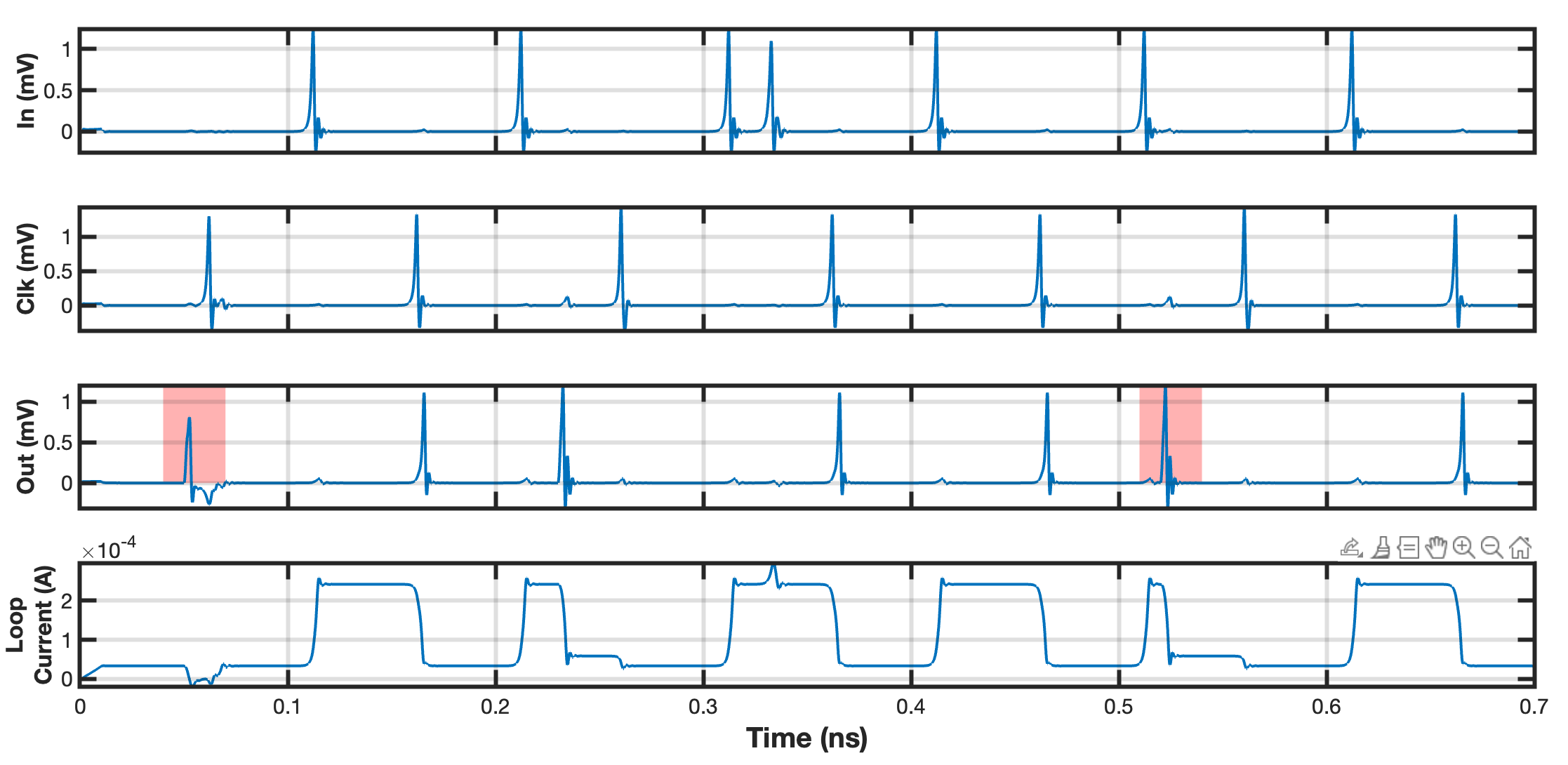}
    \centering
    \captionof{figure}{Simulation of the dff cell.}
    \centering
    \label{fig:dff_sim}

\end{minipage}%
\end{figure}

Adding synchronous reset to the DFF will increase the complexity of the cell. In this concept, we can use an $\alpha$ cell to provide an asynchronous reset at the output port. The state machine of the DFF with $\alpha$ reset functionality is given in Fig.~\ref{fig:dff_sm}. The $\alpha$ circuit not only omits the output at the DFF but also resets the input, turning the state to the idle state. The circuit schematic remains the same as the conventional DFF gate, as given in Fig.~\ref{fig:dff_sch}. We provide the functionality of DFF with $\alpha$ cell in Fig.~\ref{fig:dff_sim}. As we can see, when we are in a state idle, the reset input keeps the state idle. However, when we are in the D state, the reset input converts the state to idle, and the arriving clk doesn't produce any output. Also, the addition of $\alpha$ cells doesn't affect the functionality of the conventional DFF cell. The circuit parameters and margin values are provided in Fig.~\ref{fig:dff_sch}.

\subsection{Toggle Flip Flop (TFF)}
Toggle Flip-Flop is an essential component in RSFQ circuits, primarily serving as a frequency divider. The TFF operates by toggling its output state with each incoming pulse on its input. Specifically, when two input pulses are applied to the TFF, a single output pulse is produced, effectively halving the frequency of the input signal. 

\begin{figure}[htb!]
\centering
\noindent\begin{minipage}{\linewidth}
    \includegraphics[width=0.7\textwidth]{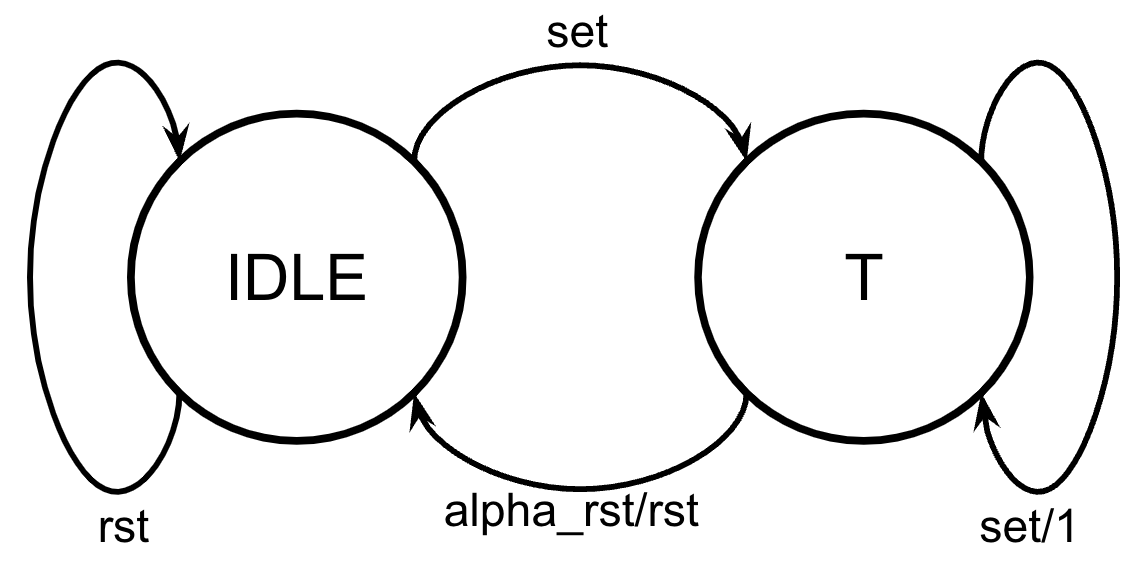}
    \centering
    \captionof{figure}{State machine of the TFF gate.}
    \centering
    \label{fig:tff_sm}
    
    \includegraphics[width=1\textwidth]{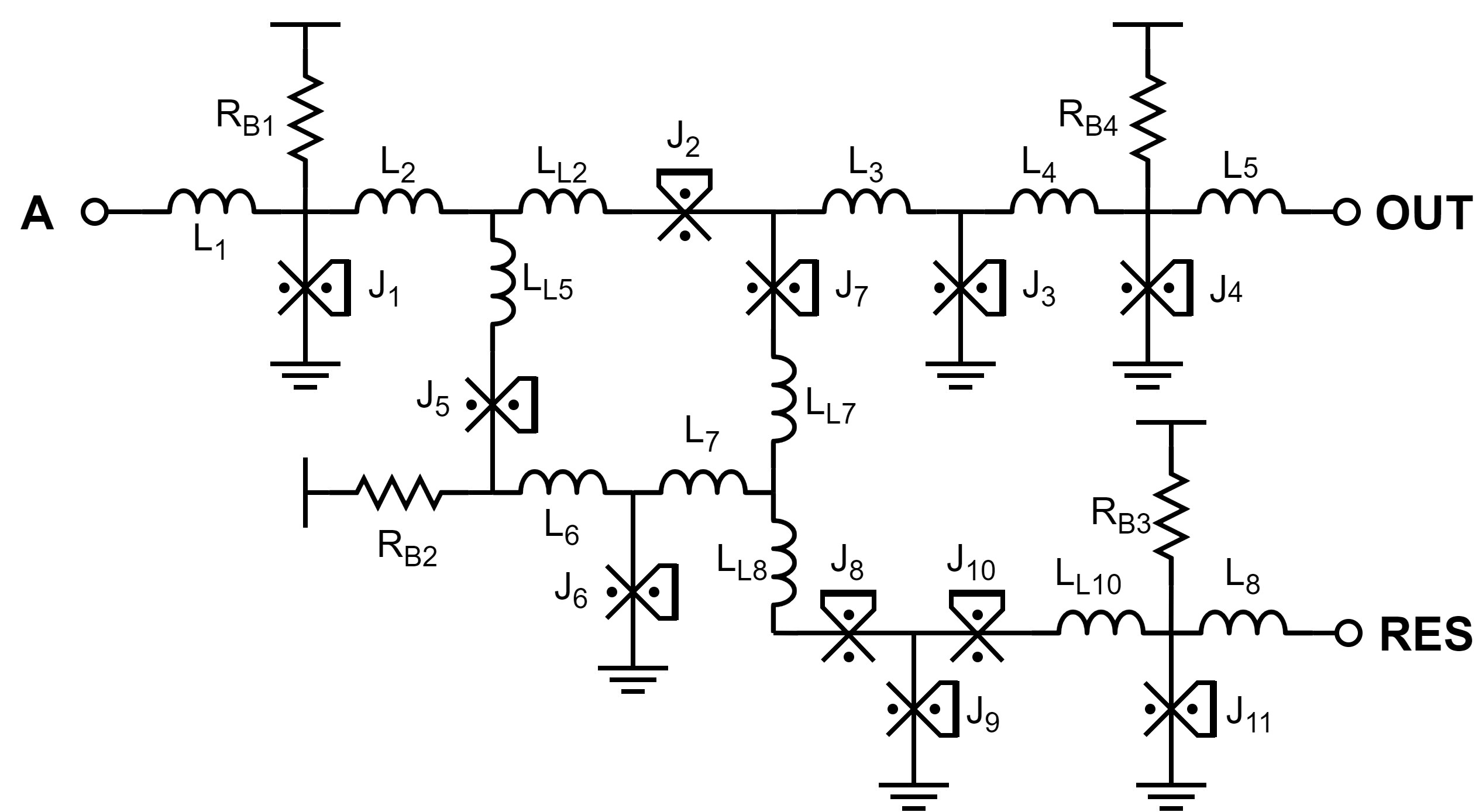}
    \centering
    \captionof{figure}{\small Schematic of the TFF gate. Configuration: L1=2.09 pH, L2=2.19 pH, LL2=1.11 pH, L3=1.51 pH, L4=5.45 pH, L5=2.23 pH, LL5=2.11 pH, L6=1.03 pH, L7=5.26 pH, LL7=1.58 pH, L8=3.25 pH, LL8=1.40 pH, LL10=2.47 pH, J1=220.08 $\mu$A, J2=124.31 $\mu$A, J3=97.05 $\mu$A, J4=180.2 $\mu$A, J5=106.24 $\mu$A, J6=102.6 $\mu$A, J7=148.8 $\mu$A, J8=100.69 $\mu$A, J9=104.31 $\mu$A, J10=104.31 $\mu$A, J11=219.13 $\mu$A}
    \centering
    \label{fig:tff_sch}
\end{minipage}

 \vspace{3 mm}
 
\begin{minipage}{\linewidth}
    \includegraphics[width=1\textwidth]{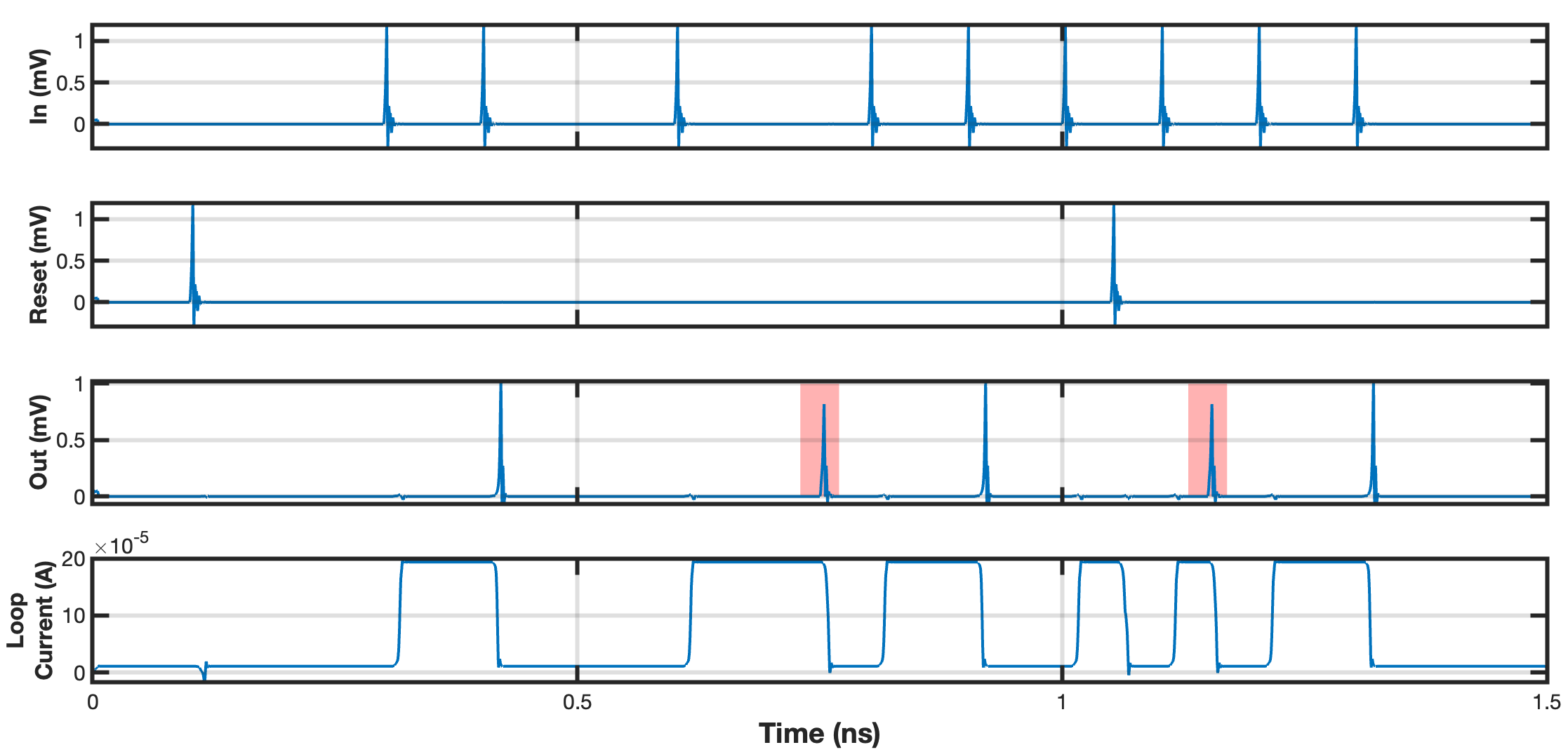}
    \centering
    \captionof{figure}{Simulation of the TFF cell.}
    \centering
    \label{fig:tff_sim}
\end{minipage}%
\end{figure}

The state machine of the Toggle Flip-Flop (TFF) is illustrated in Fig.~\ref{fig:tff_sm}. The reset input maintains the system in an idle state when the set signal is received, while the TFF transitions to the T state when the set signal is in the idle state. If another set signal arrives during the T state, an SFQ pulse is generated at the output. Conversely, if a reset signal is received while in the T state, the TFF reverts to the idle state, and the schematic representation of the Reset-Toggle Flip-Flop (RTFF) is shown in Fig.~\ref{fig:tff_sch}. However, we achieved the same functionality by routing the reset input through the output port using the $\alpha$-cell when SFQ is stored in the loop. That's why this approach allows us to eliminate the additional circuit components associated with reset functionality while retaining the reset capability through the $\alpha$-cell. Also, the $\alpha$ cell gives us the flexibility of determining the initial state of the TFF. In the simulation of the TFF, depicted in Fig.~\ref{fig:tff_sim}, we demonstrate that the reset functionality can be effectively realized using both the conventional reset port and the asynchronous reset provided via the $\alpha$-cell. The circuit parameters utilized in these simulations are detailed in Fig.~\ref{fig:tff_sch}.

\subsection{Multi-flux DRO}
\begin{figure}[htb!]
\centering
\noindent\begin{minipage}{\linewidth}
    \includegraphics[width=0.9\textwidth]{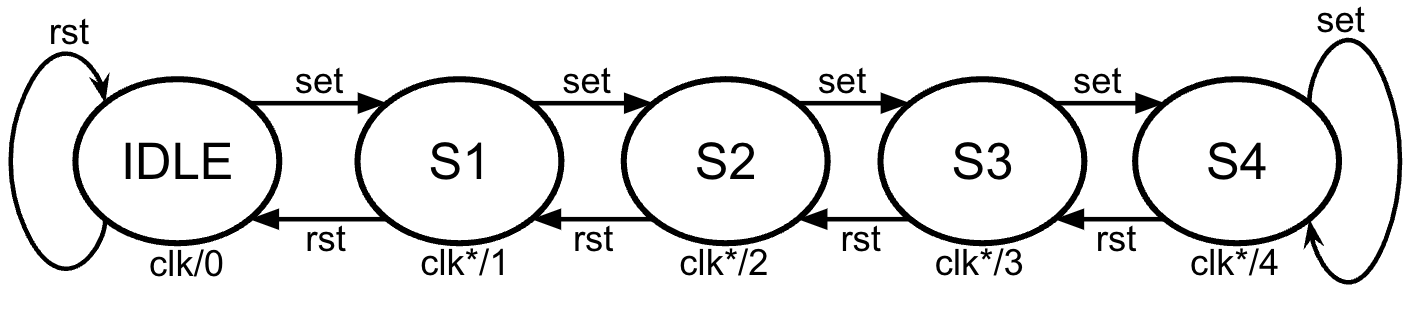}
    \centering
    \captionof{figure}{State machine of the multiflux dro.}
    \centering
    \label{fig:mdro_sm}
    
    \includegraphics[width=0.85\textwidth]{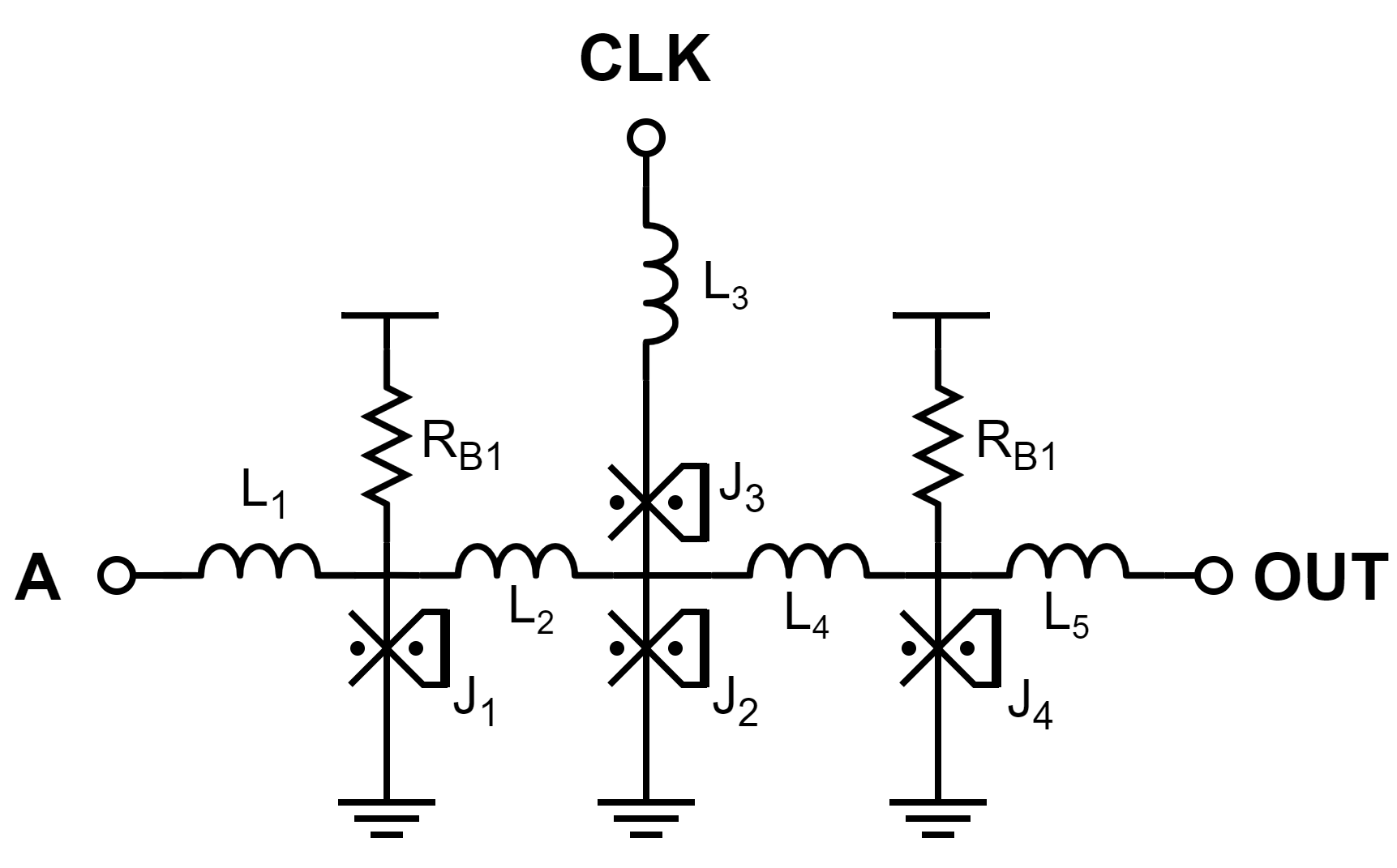}
    \centering
    \captionof{figure}{Schematic of the multiflux dro. \small Configuration: L1=0.62 pH, L2=45.31 pH, L3=2.38 pH, L4=5.52 pH, L5=2.38 pH, J1=300 $\mu$A, J2=130 $\mu$A, J3=150 $\mu$A, J4=270 $\mu$A}
    \centering
    \label{fig:mdro_sch}
\end{minipage}

 \vspace{3 mm}
 
\begin{minipage}{\linewidth}
    \includegraphics[width=1\textwidth]{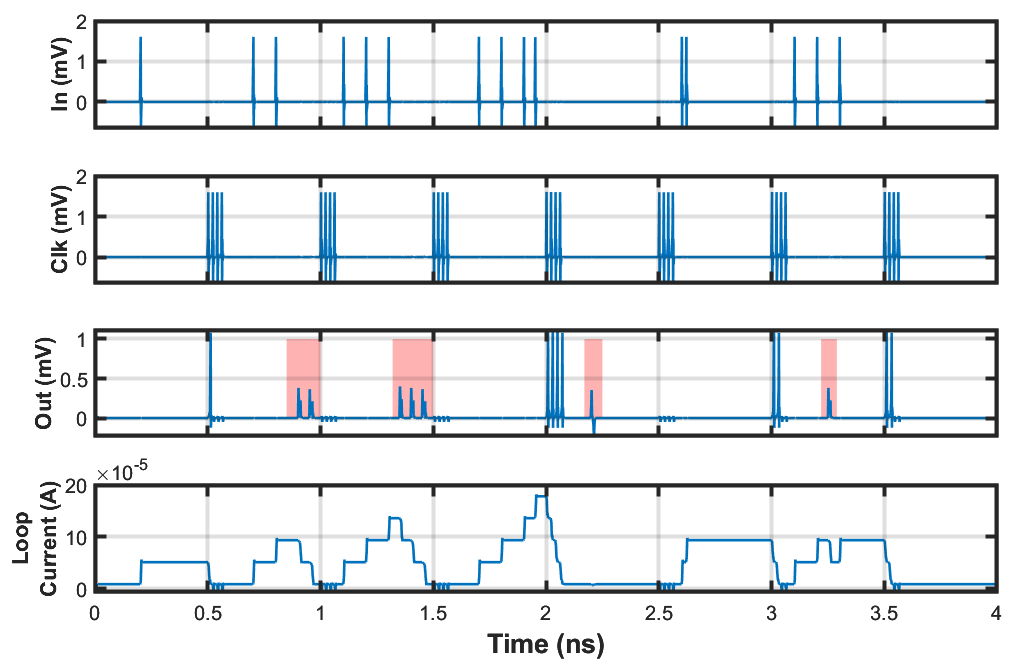}
    \centering
    \captionof{figure}{Simulation of the multiflux dro cell.}
    \centering
    \label{fig:mdro_sim}

\end{minipage}%
\end{figure}

DFF is an efficient storage unit. However, it can store only 1 bit of data, which is not enough for some cases. If we want to store multiple-bit, we need to have a multiple-bit storage unit. We can achieve a multiple-bit storage unit circuit by altering the parameters of the DFF gate. We increased the loop inductance value and junction critical current. Their multiplication gives us the storage limit. In this design, we want to store up to 4 bits, that's why \( I_c L > 4 \phi_0 \). The state machine of the circuit is given in Fig.~\ref{fig:mdro_sm}. The system has five states: idle, S1, S2, S3, S4. The input changes the state each arrival, and if clk arrives at each state, it creates 0,1,2,3,4 pulses, respectively. Meanwhile, the reset provided by the $\alpha$ cell acts as a decrement for the storage. It decreases the storage and changes the state of the system. This system can be used as an up-down counter, state machine, or any circuit that requires multiple-bit storage. 
The schematic of the multi-flux DRO is given in Fig.~\ref{fig:mdro_sch}. The structure is similar to DFF, which provides single-bit storage. The simulation result shows correct functionality with multiple-bit storage and its destruction with $\alpha$ cell given in Fig.~\ref{fig:mdro_sim}. $\alpha$ cell input decreases the storage. The circuit parameters of the multi-bit storage dro are given in Fig.~\ref{fig:mdro_sch}..

\subsubsection*{Negative pulse storage}
\begin{figure}[htb!]
    \includegraphics[width=0.5\textwidth]{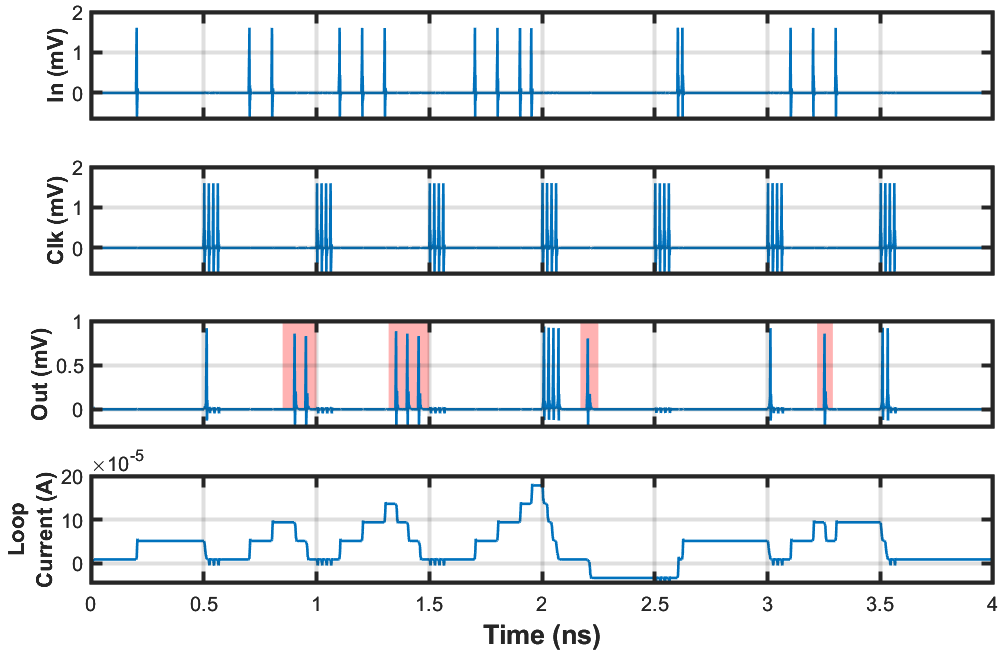}
    \centering
    \captionof{figure}{Simulation of the multiflux DRO with the negative loop.}
    \centering
    \label{fig:mdro_r_sim}
 \end{figure}

The multi-flux DRO cell, as depicted in Fig. \ref{fig:mdro_sch}, enables data storage in the negative loop by modifying its parameters. In the simulation shown in Fig. \ref{fig:mdro_r_sim}, applying an $\alpha$ input when no prior data is present results in the storage of this input as a negative flux in the loop; this mechanism grants significant flexibility, allowing the introduction of bias values into the storage loop. Once negative flux is stored, feeding two SFQ pulses through the input results in a single output pulse after clocking, with the negative pulse acting as a bias to perform a \(-1 + 2 = 1\) operation. Such functionality supports various applications, including the development of bidirectional up-down counters and other bias-related adjustments.

\section{Architectural benefit discussion}
In the shift-register-based memory architectures, specific bits must be propagated to clear targeted addresses. This process typically requires multiple clock cycles, as data must be shifted sequentially across the memory to rewrite back the rest of the values in each address. To mitigate this challenge, integrating $\alpha$-cells within DFFs enables localized resets of designated memory addresses. This approach minimizes clocking requirements and reduces the need for repetitive cycling, thereby improving both efficiency and performance in memory management.

In deep-pipelined RSFQ circuits, the insertion of targeted reset points within logic blocks significantly enhances overall efficiency. A relevant example is branch prediction: when a misprediction occurs, preliminary results generated during computation must be cleared to maintain functional correctness. By implementing local resets at specific computational stages, each logic gate can independently reset, eliminating the need for global signals and preventing the propagation of erroneous signals. This approach allows for precise control at intermediate stages. Furthermore, structures with sequential loops particularly benefit from this technique, as local resets facilitate a quicker recovery to an initial state and prevent the propagation of unnecessary signals that introduce additional delay, as these signals often cause latency while waiting for their arrival to clear the state. In these scenarios, local resets enable faster response times and reduced latency, as paths can be reset immediately after use, avoiding the need to wait for global reset signals or additional clock cycles.

\begin{figure}[htb!]
    \includegraphics[width=0.42\textwidth]{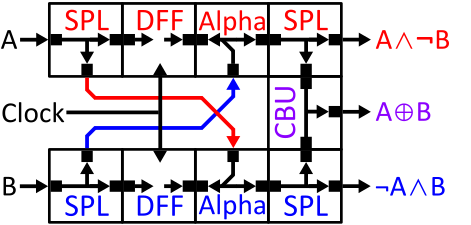}
    \centering
    \captionof{figure}{The design of XOR operation with intermediate outputs based on $\upalpha$-DFF combinations.}
    \centering
    \label{fig:XOR_3out}
 \end{figure}

As discussed in previous sections, the combination of an $\upalpha$-cell with another logic cell introduces a new input that produces the inverse logic of the output pin. For instance, consider the $\upalpha$-DFF combination: when input \( \text{A} \) is combined with the output pin \( \text{B} \) (coming from the $\upalpha$-cell), the resulting operation is \( \text{A} \land \neg \text{B} \). If the order of the pins is reversed, the operation becomes \( \text{B} \land \neg \text{A} \). By merging these two outputs using a single CBU cell, we obtain the \( \text{A} \oplus \text{B} \) operation. Notably, intermediate outputs can still be generated from both \( \text{A} \land \neg \text{B} \) and \( \text{B} \land \neg \text{A} \), in addition to their combination. Functionally, this setup generates three distinct functions across three different output pins, utilizing the same two DFFs and two $\upalpha$ cells. The related design is shown in Fig. \ref{fig:XOR_3out}.

Such a circuit provides substantial benefits in terms of resource efficiency and performance. By performing operations such as AND with an inverted logic pin and XOR using the same set of components, it minimizes hardware requirements, reduces computational complexity, and optimizes the area. The ability to implement multiple logical functions with fewer components not only enhances scalability but also contributes to a more compact and energy-efficient design in the overall architecture.

\section{Conclusion}

This work presents an asynchronous reset RSFQ (AR-RSFQ) cell library, developed without increasing the overall area and bias overhead of individual cells. The timing constraints and functional characteristics of both logic and memory cells are examined, highlighting potential extensions for their applications. The correct functionality across all possible scenarios were tested via analog simulations, and parameter values were documented. Incorporation of the reset as a third input to standard logic gates is discussed, as a distinguishing feature. Each gate can now execute a reset-modified function, described as \( \neg \text{RES} \land  \text{f(A,B)} \) where f(A,B) represents the original logic operation. A more compact and robust design of the inverter cell was developed, enhancing the library's utility. The asynchronous reset will allow for more compact and novel architecture design in the SFQ circuits. 

\bibliographystyle{IEEEtran}
\bibliography{references}

\end{document}